# Structure and Growth in the Living Tissue and in Carbon Nanotubes

Michael Pyshnov[1] and Sergei Fedorov[2]

[1]*Toronto, Canada;* [2]*Moscow, Russian Federation*

The topological organisation of cells in a model of living tissue (the crypt of intestinal epithelium) is identical to the topological organisation of atoms in carbon nanotubes. The existing models of growth of these two structures contain identical elements. It is proposed that the growth of carbon nanotubes can depend also on other mechanisms postulated in the crypt model where the growth in the bottom is transmitted to the cylinder, and, in this case, the growth of nanotubes should involve the same peculiar lattice transformations as the ones found in the crypt models. The crypt models also suggest a possible initiation of growth in the nanotubes by the loss of carbon atoms.

We consider the crystalline structures that can be formed from graphene and the structures of the living tissues as one distinct class of structures.

## 1. Introduction

The structure of tissues in the organisms emerges as a result of division of cells during the process of development. In the emerging tissues, cells are physically attached to each other in the order and at the place in which they appeared after cell division. To the every curious microscopist, tissues appear as ordered, even crystalline structures, although, with a complex symmetry. The tissue structure, however, presents a greater puzzle: in many tissues, the cells continue proliferation in the adult organism while these tissues and organs maintain their size and shape unchanged and the cells remain connected with each other.

The first attempt to solve this puzzle was made with topological modeling of cell division in the crypt of intestinal epithelium [1]. The intestinal crypt is one of the tissues that consist of one layer of cells and can topologically be considered as a two-dimensional curved and folded lattice. In the crypt, the lattice forms a cylinder closed at one end (Fig. 1 and Fig. 2); such shape is often observed in other tissues, in plants and animals. In 2005, we published a computer simulation demonstrating the process of cell division in crypt [2].

It then appeared that the topological structure in the crypt model was identical with the later found structure of carbon nanotubes. The similarity was noted in the work on graphene where the structure of the crypt was called "a living curiosity" [3].

Furthermore, while the living tissues grow by proliferation of cells within the tissue (interstitial growth), carbon nanotubes as well as other graphene structures, similarly, are capable to grow by absorbing new carbon atoms (in this case, from the surrounding medium) and intercalating them between the elements of the already formed lattice.

Here, we will compare topological structures of the crypt and the carbon nanotubes (CNT) and consider possible mechanisms of interstitial growth in these structures.

## 2. The model

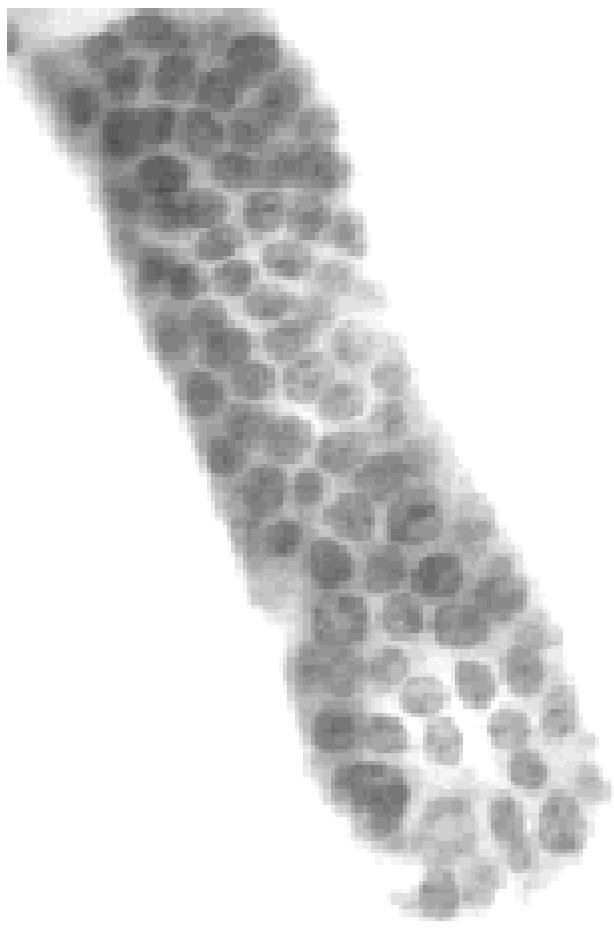

Fig. 1. Photo of the crypt of intestinal epithelium, from [2]. Cell niclei are stained.

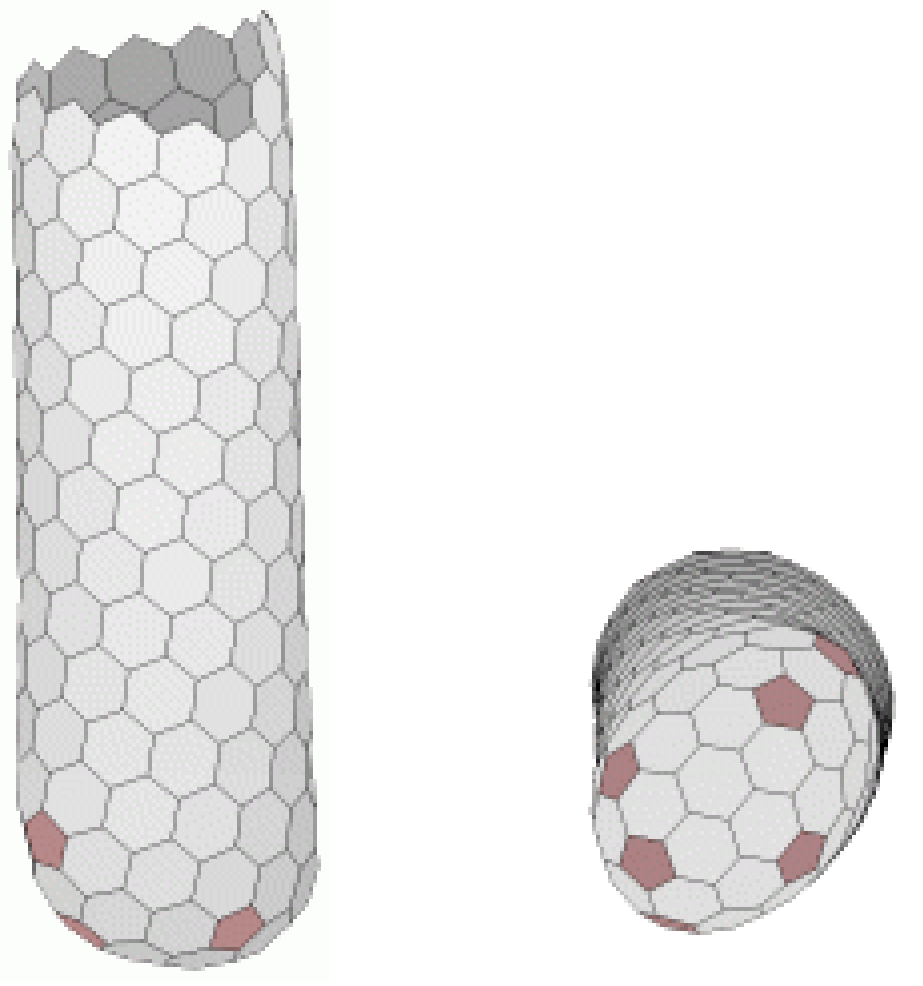

Fig. 2. The crypt model [1], side view and the bottom. Spaces inside the polygons are cells. The model turns into a carbon nanotube if vertices are atoms of carbon and the sides of the polygons are chemical bonds. Pentagons are red.

In the model, polygons of the bottom are fully connected to the polygons of the cylinder, so, the bottom represents a topological closure in which there are no unattached cell sides at one end of the crypt cylinder. In the CNT, the closure eliminates dangling chemical bonds at one end of the CNT cylinder. The closure at the bottom is achieved by introduction of six pentagons (we call them structural pentagons) into the hexagonal lattice. The positions of the six pentagons determine the configuration of the bottom and the configuration of the cylinder.

This is an idealised model: due mainly to the variations in cell size, the actual cellular lattice is not as regular as it appears here. The crypt model, therefore, represents only the net outcome of the topological events which is dictated by the Euler's theorem. The lattice is much more regular in the CNT.

## 3. The lattice defects, the division wave and other mechanisms involved in the crypt growth

In a hexagonal lattice, polygons other than hexagons are considered topological defects (dislocations). In a two-dimensional lattice, the introduction of some topological defects (a pentagon or a heptagon) will necessitate the removal or addition of large parts of the lattice and, if the elements of the lattice cannot differ much in size - will also force it to curve and fold. However, introduction of the pair pentagon-heptagon can cause only a minor buckling on the surface of the lattice. The important property of this 5-7 pair is that it can appear to be moving within the lattice when the elements of the lattice consecutively exchange their connections (a gliding dislocation). The 5-7 defect can also move in another direction within the lattice by the addition or the loss of the elements of the lattice (a climbing dislocation).

The proposed mechanism of cell division in crypt [1] was based on the moving of the 5-7 pair along the row of hexagons, inducing the cells to divide and resulting in a division wave (Fig. 3). Such interstitial growth does not require cells to break up any existing connections.

(a) (b)

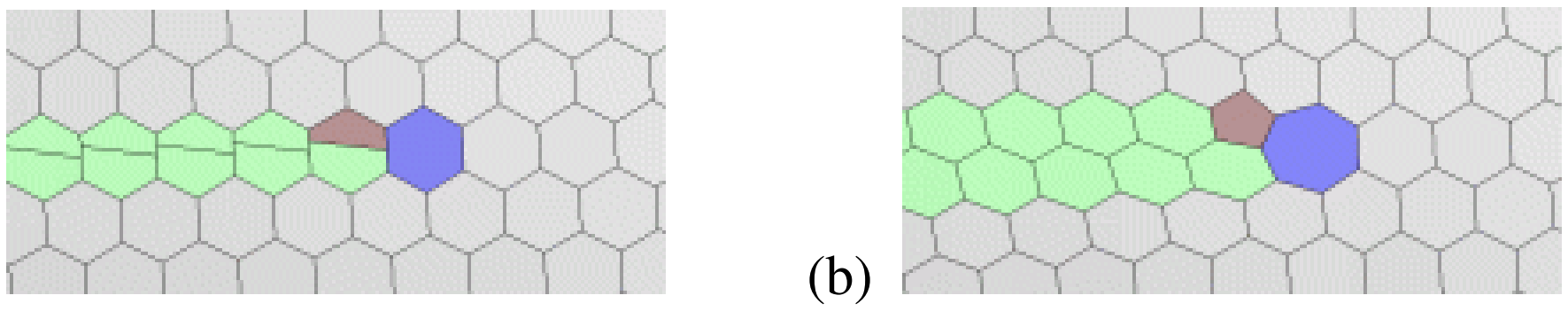

Fig. 3. The division wave proceeding from left to right. (a) - the planes of cell division are indicated; (b) - cell shape and size are partially restored. Heptagons are blue; new cells are green. Heptagon is leading the wave; in the next step it will divide, directing the plane of division to the pentagon behind (making it a hexagon) and to the hexagon ahead (making it a heptagon). Note: since there is more than one hexagon ahead, the division wave, generally speaking, can make turns.

The crucial finding in the model that allowed the crypt to grow in length while preserving its original diameter was that the division wave should first occur in the crypt bottom, in a row of hexagons situated between two pentagons (Fig. 4).

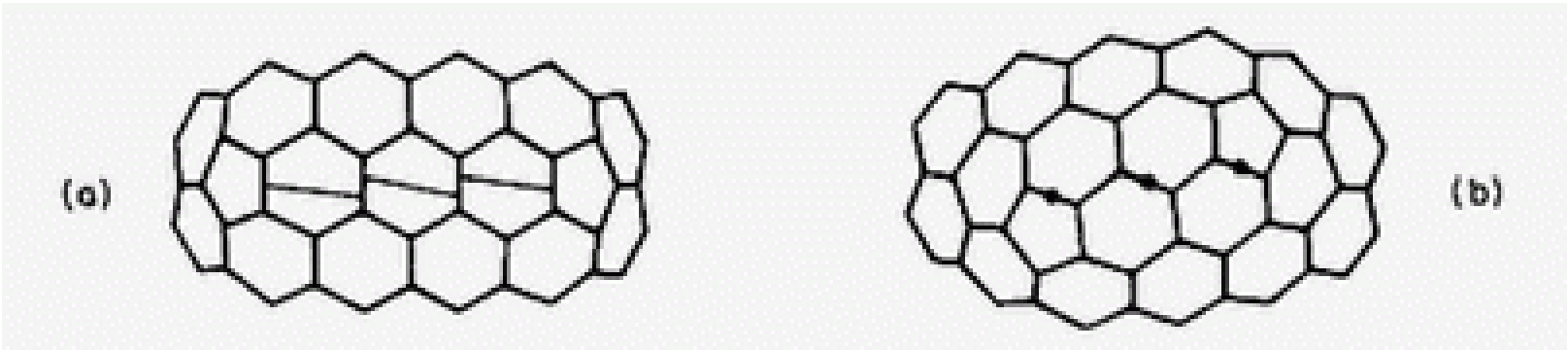

Fig. 4. The division of hexagons between two structural pentagons in the bottom, from [1]. In (a), the pentagons are on the same row with three hexagons. In (b), the hexagons have divided; the new sister cells are shown connected with dots. The pentagons are in new positions.

The cells in the Fig.4 divide locally in the bottom. However, if two division waves meet at the same pentagon in the bottom, a topological situation arises that gives birth to a new division wave that can now proceed into the crypt cylinder, Fig. 5, from [1]. In the process, the structure of the spiral rows of the hexagons in the cylinder undergoes a peculiar structural transformation described further below.

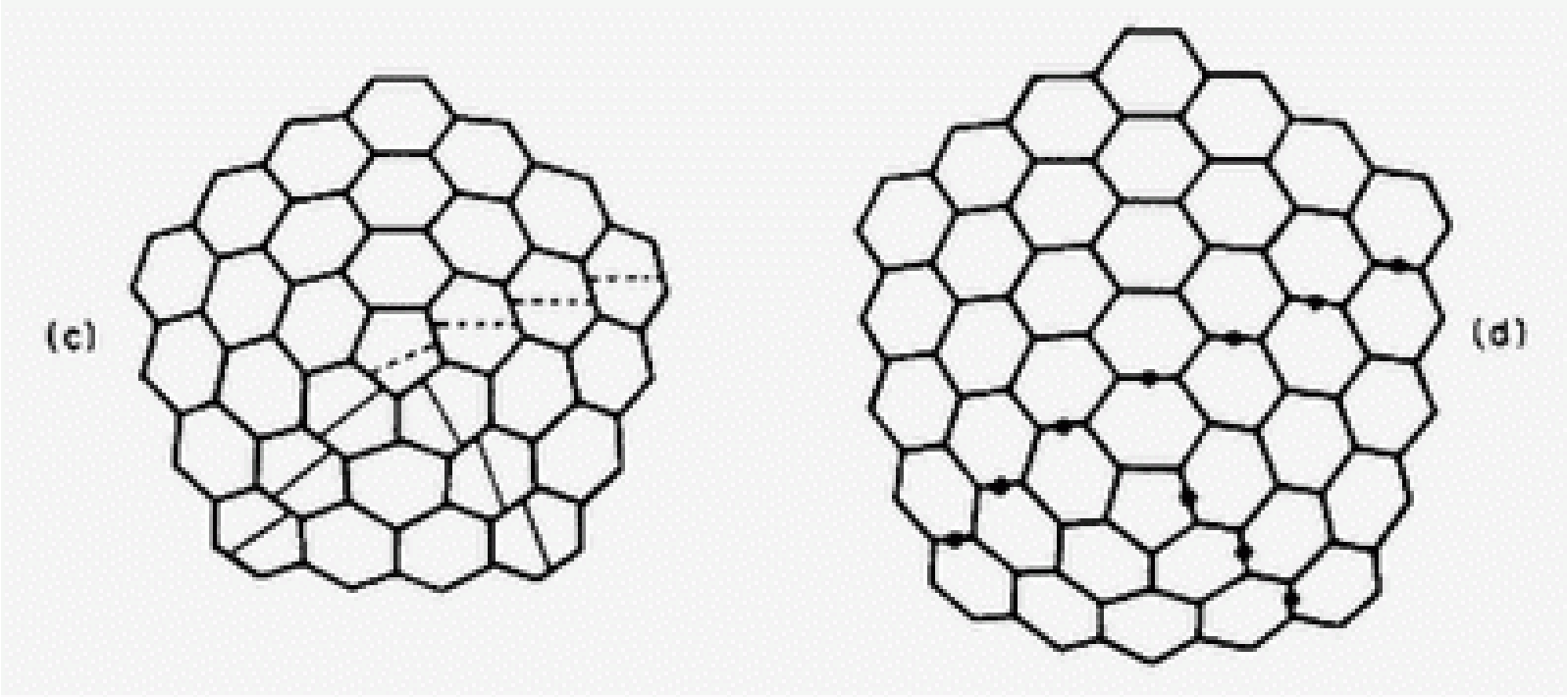

Fig. 5. (c) - two division waves meet at the pentagon and initiate a new division wave (dotted lines); (d) - cells have divided; sister cells are connected with dots.

A different mechanism for the initiation of cell division was used in [2]. It was shown that the removal/death of one pentagonal cell in the bottom topologically results in the emergence of a single 5-7 pair and in the appearance of a new pentagon restoring the required number of structural pentagons in the bottom to six (Fig. 6); see [2] for more details. The emerged 5-7 pair initiates a new division wave that can proceed in different directions. If this division wave moves to the crypt cylinder, it also results in the structural transformation of the cylinder (see below).

Let's already note here that the removal of a pentagonal cell is topologically equivalent to the loss of one $C_2$ molecule from a pentagon in the CNT. The removal of $C_2$ molecule opens 4 dangling chemical bonds that form a new configuration: two pentagons and a heptagon.

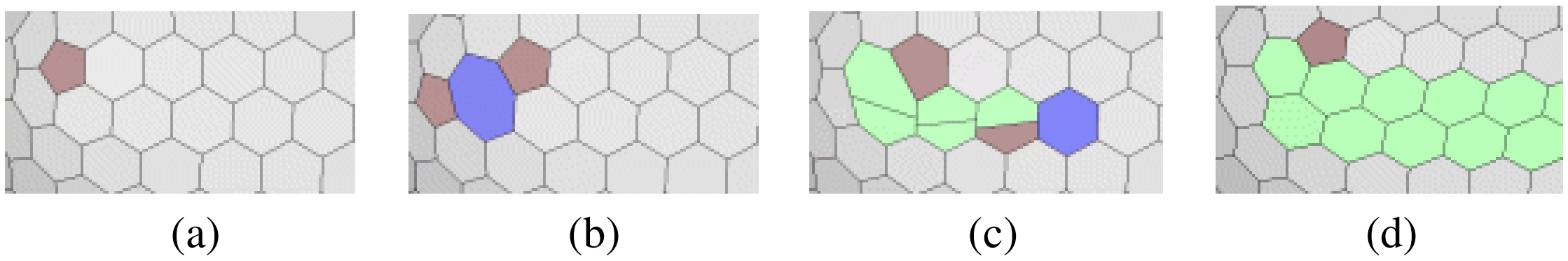

Fig. 6. The division wave initiated by the removal of a pentagon. In (b), a heptagon is formed and it is dividing in (c), directing the plane of division to one of the pentagons (here, to the pentagon that in (b) was on the left of the heptagon).

In Fig. 7, we show the same division wave as in Fig. 6, but now meeting a pentagon on the other end. This wave is proceeding between two pentagons which, however, are not on the same row of hexagons, because, as it is now obvious, the division wave initiated by the removal of a pentagon is making a turn. This case is important because such division wave follows the path known in the CNT growth models as "twisted Endo-Kroto patch" [6]. It is "twisted" since the two pentagons are not on the same row of hexagons.

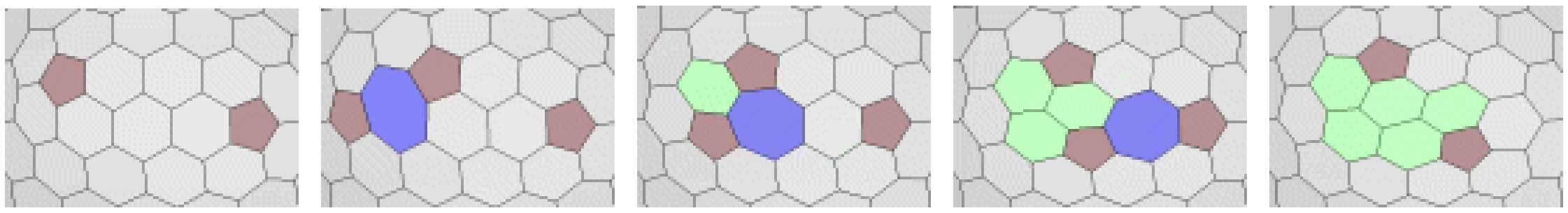

Fig. 7. The division wave initiated by the removal of pentagon in the topological situation of the twisted Endo-Kroto patch. Consecutive cell divisions are shown, from left to right.

## 4. Structural transformations in the crypt cylinder

The transformations accompanying the passing of the division wave in the crypt cylinder were described in [1]. Later, the model using the removal of a pentagon as the mechanism for initiation of cell division, with much greater choice of the outcomes, was studied with the computer animation program [2]; it showed the same structural transformations in the cylinder.

Depending on the position and orientation of the arising 5-7 pair, the division waves proceeding to the cylinder result in changes of the pitch and threadedness of the spiral rows of hexagons in the cylinder (Fig. 8). The changes have cyclical character, i. e. after several consecutive division waves, the cylinder can be returned to its initial configuration. It is possible to keep the perimeter of the cylinder essentially constant (although, strictly speaking - slightly fluctuating).

Other division waves having different direction and/or orientation of the division planes (see note to the Fig. 3) can increase the perimeter of the cylinder. However, there is no such division wave that will result in the reduction of the crypt perimeter.

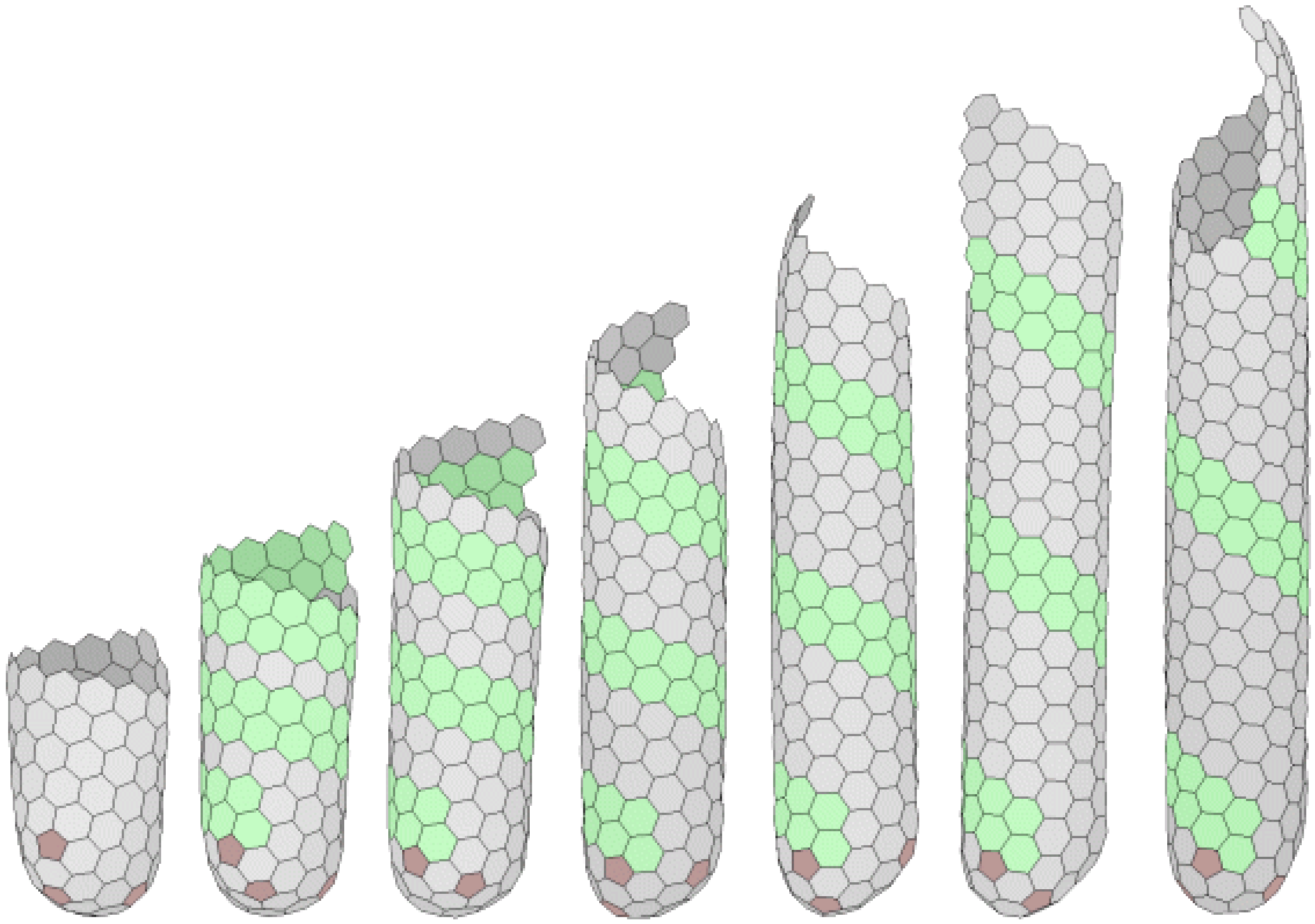

(a) (b) (c) (d) (e) (f) (g)

Fig. 8. Structural transformations in the crypt cylinder. The growth was generated by the removal of a pentagon followed by a division wave having the same orientation in all figures. Each figure was produced by the removal of the pentagon that was produced in the previous figure. This pentagon was changing its location in the bottom after every division wave. The other five pentagons of the bottom were not affected. In the CNT terminology, figures (a) to (e) are hiral structures; (f) is armchair; (g) has reverted to hiral of the same pitch as (e), but of the opposite handedness. The properties of the spiral structure (threadedness in two directions, as used in [2] and in the CNT literature) are as follows: (a) - (12,2); (b) - (11,3); (c) - (10,4); (d) - (9,5); (e) - (8,6); (f) - (7,7); (g) - (8,6).

## 5. Mechanisms of growth in carbon nanotubes

In 1992, Endo and Kroto [4] proposed that insertion of two carbon atoms ($C_2$ molecule) into a hexagon situated between two pentagons can serve as a basic mechanism of growth of the fullerenes and the CNT. The result was the "division" of the hexagon into two pentagons and the addition of one side to each original pentagon making them hexagons, i. e. the case described for the cell division in the crypt bottom and many times shown in CNT literature in various configurations and with various numbers of hexagons "dividing" between two pentagons; this being known as the Endo-Kroto patch. The authors [4] suggested that "*active sites* [for the absorption of $C_2$ molecule] *are in the vicinity of pentagons and that the process involves insertion-reorganisation process in which strain is equilibrated by diffusion of the pentagonal cusps*".

The situation around the pentagons in the crypt bottom was described similarly: "*the crypt cells are trying to eliminate dislocations and rectify the surface by means of cell division*" [1], and it was suggested that the six structural pentagons that cannot be removed from the crypt bottom and can only change their positions as a result of cell division, remain there as permanent sources of instability initiating new cell division and growth in the crypt.

Saito et all., also in 1992 [5], noted the state of electronic excitation that exists in a row of hexagons situated between two pentagons in the CNT bottom (where "*the curvature of the bonds between two pentagons is larger than for the other bonds*") and also said that "*those hexagons between the two pentagons can be considered as active sites for absorbing a $C_2$ molecule.*" They noted that when the bottom grows, the new hexagons at the perimeter of the bottom will become a part of the cylinder, producing its growth in length. The authors also described a number of topological situations, such as changing positions of the pentagons in the CNT bottom and the meeting of two dislocations at a *hexagon.* Yet, they made no suggestion that a meeting of two dislocations at a *pentagon* can result in the formation of a new 5-7 pair (as in Fig. 5).

It is now accepted that the absorption of new carbon atoms and the growth of the CNT is possible due to the strain on the chemical bonds near the pentagonal defects.

In both 1992 papers that defined the role of the pentagons in the CNT growth, the growth occurs only locally (between two pentagons in fullerenes and in the CNT bottom), as there are no pentagons in the cylinder; there is no absorption of carbon atoms in the cylinder.

Other models of CNT growth describe the growth from the open end of the cylinder (strictly speaking, not an interstitial growth). These are interpreting the results of an experimental procedure that allows CNT growth on a grain of a catalyst attached to the CNT open end. The essential difference of the open end growth models is that they do not explicitly use pentagonal lattice defects as active sites for the absorption of $C_2$ molecules; the open end itself (while being in contact with catalyst) represents the active site. Here, the topological situation allowing the growth of the CNT remains obscure. Yet, one interesting model suggests the topological arrangements that should allow the CNT cylinder to grow as a node extending from a flat sheet of graphene [7].

## 6. Are there division waves in the CNT?

The mechanisms of the CNT growth discussed above produce only a local growth at one of the ends of the cylinder. They explain the elongation of the cylinder and, in some cases, are able to grow the new part of the cylinder with a different diameter. They, however, cannot change the diameter of the already formed part of the cylinder. Yet, the typical experiments show not only axial growth, but also increase in the CNT diameters which however are seen as remarkably uniform through the entire length of the individual CNTs. Topologically, this would require the addition of new atoms along the entire length of the cylinder, but that does not seem to be possible within the proposed mechanisms. Now, the mechanisms of growth proposed in the crypt model where the division wave is proceeding along the row of hexagons in the cylinder, allow the cylinder to grow in length as well as to uniformly increase its diameter (depending on the direction of the wave and the orientation of the planes of division). We are proposing here that these mechanisms might operate in the CNT. The "division" waves in the CNT, i. e. sequential absorption of carbon atoms in a row of hexagons, can probably be responsible for maintaining the wonderful uniformity of the diameter of the CNT cylinder over its entire length as well as its uniform changes. Such mechanism can be responsible for maintaining the structure relatively free from defects.

Generally speaking, the "division" of several hexagons between two pentagons (Endo-Kroto patch with several hexagons) can be regarded as a division wave starting at one pentagon (or - at both) and self-annihilating when all hexagons have divided. The strain on the bonds is higher at the hexagons immediately adjacent to the pentagons, which makes these hexagons the likely initiation points for the absorption of $C_2$ molecules, in which case the process of absorption would represent a wave, not a simultaneous reaction in the hexagons.

Let's now remember that the chemical bonds of the hexagons on a curved surface of the CNT cylinder are under the radial strain. This strain is generally smaller than the strain near the pentagonal defects in the bottom (that of course depends on the radius of the cylinder), but, it seems possible that the radial strain can assist the absorption of the $C_2$ molecules in a row of hexagons in the cylinder once the wave has been initiated in the bottom. If the initiation of $C_2$ absorption requires the adjacency to a pentagon, the further propagation of the wave might not. This is also the case in the row with several hexagons between two pentagons where the surface curvature at the hexagons not adjacent to pentagons is similar to the curvature at the cylinder.

Furthermore, the strain on the bonds belonging to the different sides of the hexagons in the cylinder is unequal, it depends on the orientation of the hexagons in the cylinder. In the cylinder and parallel to the spiral directions, there are lines involving the bonds under greater radial strain and the lines where the strain on the bonds can be smaller or absent (that depends of the particular configuration). The $C_2$ absorption should be favored along the lines where the strain on the bonds is greater. And that probably can play a role in defining the direction of the division waves and the orientation of the division planes (see note to the Fig. 3).

If the above propositions are correct, there could be a cessation of the division waves in the cylinder and a much slower growth in the CNT having larger diameter.

If the experiments prove that the individual CNT cylinder can, during the growth, change its configuration (armchair, chiral and zigzag), this can be explained as a result analogous to the one shown in Fig. 8.

## 7. A class of structures

A number of structures can be produced from graphene: spheres, tubes, nodal growth, branching, etc. These structures closely resemble living tissues and they are also capable of interstitial growth. Interestingly, branching is observed also in the intestinal crypts. Whether produced from graphene by absorption of atoms from the surrounding medium or produced by the division of cells, these structures have common properties. While completely different in the operating physical forces, these structures are produced by the higher law, that of topology. We unite them in one class of structures. The main properties that we believe exist in this class of structures are:

1. The structures are formed by the topological laws operating only in two dimensions and that allows the lattice to produce a considerable variety of patterns.
2. Most importantly, the structures can combine in one lattice parts having different shape and symmetry. They in fact are multi-periodic crystals.
3. When, in this class of structures, dislocations are introduced, they might not lead to the weakening of the structure, but, instead, will lead to the local changes in the shape and symmetry of the contiguous lattice. Even the term *dislocation* here has a different meaning - this is not necessarily a destructive defect in the lattice because the lattice can remain fully contiguous; the result can be that only a domain with a different shape and symmetry is formed.

4. The lattice in this class of structures is capable of interstitial growth, which would be very difficult to achieve in a three-dimensional lattice.

5. While a three-dimensional lattice is always open, i. e. it allows indefinite growth of the crystal on its edges, the two-dimensional lattice can close on itself to form a topological sphere. The growth in such closed structure can only be interstitial.

6. When only the interstitial growth is possible, there should exist intrinsic topological conditions specifying the patterns of growth and limiting the size of the structure and the number of various shapes that can be produced. In living organisms, normal cells divide only within the contiguous lattice. Evolution therefore has the above limitations. There should also exist conditions determining the possibilities and impossibilities of the process of regeneration of the lost parts.

7. A multicellular organism developing from an embryo to a certain adult shape and size can essentially be, at all stages, a topological sphere.

## Notes

1. In this paper, we are concerned with the climbing dislocations producing growth, i. e. the continual addition of carbon atoms capable of preserving the regular structure of the CNT. It is well known that structural transformations (the changes in the CNT diameter and length, change of handedness, etc.) can also be caused by the gliding dislocations, by the climbing dislocations that move by removal of the lattice elements and by other defects acting locally. On the other side, the gliding dislocations cannot be considered in the living tissue if the cell-cell contacts are permanent.

2. The important factor considered in suggesting different mechanisms of the CNT growth is which mechanism of the $C_2$ absorption is energetically more favored. There are indications that the division wave (which is a 5-7 climbing dislocation moving by the addition of lattice elements) should be energetically costly. However, the actual mechanism depends on the choice of the experimental procedure. The procedure, temperature in particular, can vary greatly, but, apparently, at higher temperatures, different reactions are possible at the same time, producing the regular structure as well as different topological defects. We wonder if the way to produce a regular structure could possibly be to reduce the temperature and the variety of fragments available in vapor and to supply larger fullerene fragments that can act as donors of $C_2$ molecules by breaking up and releasing $C_2$ upon the contact with the specific sites on the CNT.

3. Images with color in the article were generated with the 3-D computer simulation program [2].